\documentstyle[epsf,psfig,12pt]{article}
\renewcommand{\baselinestretch}{1.05}

\begin{document}
\def\be{\begin{eqnarray}}
\def\en{\end{eqnarray}}
\def\up{\uparrow}
\def\dw{\downarrow}
\def\non{\nonumber}
\def\la{\langle}
\def\ra{\rangle}
\def\nc{N_c^{\rm eff}}
\def\ep{\varepsilon}
\def\vp{\varepsilon}
\def\vma{{_{V-A}}}
\def\vpa{{_{V+A}}}
\def\m{\hat{m}}
\def\fp{{f_{\eta'}^{(\bar cc)}}}
\def\half{{{1\over 2}}}
\def\pr{{\sl Phys. Rev.}~}
\def\prl{{\sl Phys. Rev. Lett.}~}
\def\pl{{\sl Phys. Lett.}~}
\def\np{{\sl Nucl. Phys.}~}
\def\zp{{\sl Z. Phys.}~}
\font\el=cmbx10 scaled \magstep2
{\obeylines
\hfill NCKU-TH-98-04
\hfill July, 1998}

\vskip 1.5 cm

\centerline{\large\bf Exclusive charmless $B_s$ hadronic decays}
\centerline{\large\bf into  $\eta'$ and $\eta$ }
\medskip
\bigskip
\medskip
\centerline{\bf B. Tseng}
\medskip
\bigskip
\centerline{Department of Physics, National Cheng-Kung University}
\centerline{Tainan, Taiwan 700, Republic of China}
\bigskip
\centerline{\bf Abstract}
\bigskip
{\small   
Using the next-to-leading order QCD-corrected effective Hamiltonian, 
charmless exclusive nonleptonic decays of the $B_s$ meson  into  
$\eta$ 
or $\eta'$ are calculated within the generalized factorization approach.
Nonfactorizable contributions, which can be parametrized in terms of  
the effective 
number of colors $N_c^{\rm eff}$ for $P P$ and $V P$ decay modes, are
studied in two different schemes: 
(i) the one with the ``homogeneous" structure in which 
 $(N_c^{\rm eff})_1 \approx (N_c^{\rm eff})_2 \approx \cdots \approx 
(N_c^{\rm eff})_{10}$ is assumed, and 
(ii) the ``heterogeneous" one in which  the possibility of 
$N_c^{\rm eff}(V+A) \neq N_c^{\rm eff}(V-A)$ is considered, where 
$N_c^{\rm eff}(V+A)$ denotes the effective value of colors for  the 
$(V-A)(V+A)$ penguin
operators and $N_c^{\rm eff}(V-A)$ for
the $(V-A)(V-A)$  ones.
For processes depending  on the $N_c^{\rm eff}$-stable $a_i$ 
such as $\bar{B_s} \to (\pi,\rho) (\eta^{'},\eta)$,
the predicted branching ratios are not sensitive to the factorization 
approach we choose. While for the processes depending  on the 
$N_c^{\rm eff}$-sensitive 
$a_i$ such as $\bar{B_s} \to \omega \eta^{(')}$, there is a 
wide range 
for the  branching ratios depending on the choice of the $N_c^{\rm eff}$
involved.
We have included the QCD anomaly effect in our calculations  
and  found that it is important for $\bar{B_s} \to \eta^{(')} 
\eta^{(')}$. 
The effect of the  $(c\bar c) \to \eta'$ mechanism is found to 
be tiny  due to a possible CKM-suppression and the suppression 
in the  decay constants
except for  the $\bar{B_s} \to \phi \eta$ decay within the ``naive"
factorization approach, where
the internal $W$ diagram is CKM-suppressed and
the penguin contributions are compensated.
}

\pagebreak
 
{\bf 1.}~{\it Motivation}~~Stimulated by the recent observations of 
the large inclusive and exclusive rare $B$ decays by 
the CLEO Collaboration \cite{CLEO}, there are considerable 
interests in the charmless $ B $ meson decays \cite{Chau1}. 
To explain the abnormally large  branching ratio of
the semi-inclusive process $B\to\eta'+X$,
several mechanisms have been advocated \cite{AS,HT,Frit,mechanism}  and 
some tests of these mechanisms have been proposed \cite{DuYang97}.
It is now generally believed that the QCD anomaly \cite{AS,HT,Frit} 
plays a vital role.
The understanding of the exclusive $B \to \eta' K$, however, relies on 
several subtle points. First, the QCD anomaly does occur through the
equation of motion \cite{KP97,Ali97} when  calculating the $(S-P)(S+P)$ 
penguin 
operator and its effect is found to reduce the branching ratio.
Second, the mechanism of $c\bar{c} \to\eta'$,
although proposed to be large and positive originally \cite{HZ97,SZ98}, 
is now preferred to be negative and smaller than before as implied by 
a recent theoretical recalculation \cite{AMT} and several phenomenological 
analyses \cite{Ali97,FK97}. Third, the running strange quark mass 
which appears in the calculation of the matrix elements of the $(S-P)(S+P)$ 
penguin operator, 
the $SU(3)$ breaking effect in the involved $\eta'$ decay constants 
and the normalization of the $B \to \eta^{(')}$ matrix element involved 
raise the branching ratio substantially.
Finally, nonfactorizable contributions, which are parametrized by 
the $N_c^{\rm eff}$, gives the final answer for the largeness of exclusive
$B \to \eta' K$. ( We refer the reader to \cite{CT98,ALI98} for 
details.)

It is very interesting to see the impacts of these subtleties mentioned
above on the the exclusive charmless $B_s$ decays to an
$\eta'$ or $\eta$. In addition to  the essential and important QCD
penguin contribution as discussed in \cite{DX,DBGFN}, 
it is found that the EW penguin contribution is important for some 
processes \cite{EW}, $\it e.g.$ $B_s \to (\pi,\rho) (\eta,\phi)$
which the QCD penguin does not contribute to.
However, the effects of QCD anamoly on the $B_s$ decay 
 are not discussed in  earlier papers. 
This motivates us to consider 
the contributions of anamoly effects in charmless $B_s$ decays. 
Another interesting topic we would like to study is the importance of 
the mechanism $c\bar{c} \to\eta'$. 
Besides, the running quark  mass, the $\eta^{(')}$
decay constant and the normalization of the matrix element involving 
$\eta^{(')}$ are carefully taken care of in this Letter.
  
\medskip
{\bf 2.}~{\it Theoretical Framework}~~We begin with a brief description of 
the theoretical framework. The relevant effective $\Delta B=1$ 
weak Hamiltonian is
\be
{\cal H}_{\rm eff}(\Delta B=1) = {G_F\over\sqrt{2}}\Big[ V_{ub}V_{uq}^*(c_1
O_1^u+c_2O_2^u)+V_{cb}V_{cq}^*(c_1O_1^c+c_2O_2^c)
-V_{tb}V_{tq}^*\sum^{10}_{i=3}c_iO_i\Big]+{\rm h.c.},
\en
where $q=d,s$, and
\be
&& O_1^u= (\bar ub)_\vma(\bar qu)_\vma, \qquad\qquad\quad~~O_1^c = (\bar cb)_
\vma(\bar qc)_\vma,   \non \\
&& O_2^u = (\bar qb)_\vma(\bar uu)_\vma, \qquad \qquad \quad~~O_2^c = 
(\bar qb)_\vma(\bar cc)_\vma,   \non \\
&& O_{3(5)}=(\bar qb)_\vma\sum_{q'}(\bar q'q')_{\vma(\vpa)},  \qquad
O_{4(6)}=(\bar q_
\alpha b_\beta)_\vma\sum_{q'}(\bar q'_\beta q'_\alpha)_{\vma(\vpa)},
\non \\
&& O_{7(9)}={3\over 2}(\bar qb)_\vma\sum_{q'}e_{q'}(\bar
q'q')_{\vpa(\vma)},
\quad
O_{8(10)}={3\over 2}(\bar q_\alpha b_\beta)_\vma\sum_{q'}e_{q'}(\bar
q'_\beta
q'_\alpha)_{\vpa(\vma)},   
\en
with $(\bar q_1q_2)_{_{V\pm A}}\equiv\bar q_1\gamma_\mu(1\pm \gamma_5)q_2$. 
In Eq.~(2), $O_{3-6}$ are QCD penguin operators and $O_{7-10}$ 
are electroweak penguin operators. 
$C_i(\mu)$ are the Wilson coefficients, which have been evaluated  
to the next-to-leading order (NLO) \cite{Buras92,Ciuchini}. 
One important feature of the NLO calculation is 
the renormalization-scheme and -scale dependence of 
the Wilson coefficients 
(for a review, see \cite{Buras96}). 
In order to ensure the $\mu$ and renormalization scheme independence for 
the physical amplitude, the matrix elements, which are
evaluated under the factorization hypothesis, 
have to be computed in the same renormalization scheme and 
renormalized at
the same scale as $c_i(\mu)$. 
However, as emphasized in \cite{CT98},  the
matrix element $\la O\ra_{\rm fact}$ is scale independent under the
factorization approach and hence it cannot be identified with $\la
O(\mu)\ra$. 
Incorporating  QCD
and electroweak corrections to the four-quark operators, we can redefine
$c_i(\mu)\la O_i(\mu)\ra= {c}_i^{\rm eff}\la O_i\ra_{\rm tree}$,
so that ${c}_i^{\rm eff}$ are  renormaliztion scheme 
and scale independent.
Then the factorization approximation is applied to the hadronic matrix
elements of the operator $O$ at the tree level. 
The numerical values for ${c}_i^{\rm eff}$ are 
shown in  the last column of Table I, where $\mu={m}_b(m_b)$,
$\Lambda^{(5)}_{\overline{\rm MS}}=225$ MeV, $m_t=170$ GeV and
$k^2=m_b^2/2$ are used \cite{CT98}.

In  general, there are contributions from the nonfactorizable
amplitudes. 
Because there is only one single form factor (or Lorentz scalar)
involved in the decay amplitude of $B\,(D)\to PP,~PV$ decays ($P$:
pseudoscalar meson, $V$:
vector meson), the effects of nonfactorization can be lumped into the
effective parameters $a_i^{\rm eff}$ \cite{Cheng}:
\be
a_{2i}^{\rm eff}=c_{2i}^{\rm eff}+c_{2i-1}^{\rm eff}\left({1\over
N_c}+\chi_{2i} \right),\qquad 
a_{2i-1}^{\rm eff}=c_{2i-1}^{\rm eff}+c_{2i}^{\rm eff}\left({1\over
N_c}+\chi_{2i-1}\right),
\en
where $c_{2i,2i-1}^{\rm eff}$ are the Wilson coefficients of 
the 4-quark operators,
and nonfactorizable contributions are characterized by the parameters
$\chi_{2i}$ and $\chi_{2i-1}$.
We can parametrize the nonfactorizable contributions by 
defining an effective number of colors $N_c^{\rm eff}$, called $1/\xi$ in
\cite{BSW}, as $ 1/N_c^{\rm eff} \equiv (1/N_c)+\chi$.
Different factorization approach used in the literature 
can be classified by the effective number of colors $N_c^{\rm eff}$.
The so-called ``naive" factorization discards all the nonfactorizable
contributions and takes $ 1/N_c^{\rm eff}= 1/N_c=1/3 $, whereas
the ``large-$N_c$ improved" factorization \cite{Buras} drops out all 
the subleading $1/N_c$ terms and takes $ 1/N_c^{\rm eff}=0$.
In principle,
$N_c^{\rm eff}$ can vary from channel to channel, as in the case of charm   
decay. However, in the energetic two-body $B$ decays, $\nc$
is expected to be process insensitive as supported by data
\cite{Neubert}. If $N_c^{\rm eff}$ is process
independent, then we have a generalized factorization.  
In this paper, we will treat the nonfactorizable contributions
with two different phenomenological ways
: (i) the one with ``homogenous" structure, which 
assumes that  $(N_c^{\rm eff})_1  \approx (N_c^{\rm eff})_2 
\approx \cdots \approx (N_c^{\rm eff})_{10}$
, and (ii) the ``heterogeneous" one, 
which considers the possibility of 
$N_c^{\rm eff}(V+A)\neq N_c^{\rm eff}(V-A)$. The consideration of the
``homogenous" nonfactorizable contributions, which is commonly used in 
the literature, has its advantage of simplicity. 
However, as argued 
in \cite{CT98}, due to the different Dirac structure of the Fierz 
transformation, nonfactorizable effects in the matrix   
elements of $(V-A)(V+A)$ operators are {\it a priori} different 
from that
of $(V-A)(V-A)$ operators, i.e. $\chi(V+A)\neq \chi(V-A)$. Since
$1/N_c^{\rm eff}=1/N_c+\chi$ 
, theoretically it is expected that
\be
&& N_c^{\rm eff}(V-A)\equiv
\left(N_c^{\rm eff}\right)_1\approx\left(N_c^{\rm eff}\right)_2\approx
\left(N_c^{\rm eff}\right)_3\approx\left(N_c^{\rm eff}\right)_4\approx
\left(N_c^{\rm eff}\right)_9\approx
\left(N_c^{\rm eff}\right)_{10},   \non\\
&& N_c^{\rm eff}(V+A)\equiv 
\left(N_c^{\rm eff}\right)_5\approx\left(N_c^{\rm eff}\right)_6\approx
\left(N_c^{\rm eff}\right)_7\approx
\left(N_c^{\rm eff}\right)_8.
\en
To illustrate the effect of the nonfactorizable contribution, we
extrapolate  $N_c(V-A) \approx 2$ from 
$B \to D \pi(\rho)$ \cite{CT95}
to charmless decays. 
The $N_c^{\rm eff}$-dependence of the effective parameters $a_i$'s 
are shown in Table I, from which  we  
see that $a_1, a_4, a_6$ and $a_9$ are $N_c^{\rm eff}$-stable, and the 
remaining ones 
are $N_c^{\rm eff}$-sensitive. 
We would like to remark that while $a_3$ and $a_5$ 
are both $N_c^{\rm eff}$-sensitive, the combination of 
($a_3 - a_5$) is 
rather stable under the variation of the $N_c^{\rm eff}$ within the
``homogeneous" picture and is still sensitive to the factorization
approach taken in the ``heterogeneous" scheme. This is the main
difference between the ``homogeneous"  and ``heterogeneous"
approaches.
While $a_7,a_8$ can be neglected, $a_3,a_5$ and $a_{10}$ have some effects 
on the relevant processes depending on the choice of $N_c^{\rm eff}$.

\begin{table}
\begin{center}
{{\small Table I. Numerical values of effective coefficients $a_i$ 
at $N_c^{\rm eff}=2,3,5,\infty$, where $N_c^{\rm eff}=\infty$
corresponds to $a_i^{\rm eff}=c_i^{\rm eff}$. 
The entries for $a_3$,...,$a_{10}$
have to be multiplied with $10^{-4}$.}

\medskip
\begin{tabular}{|c|c|c|c|c|}
\hline
 & $N_c^{\rm eff}=2$ & $N_c^{\rm eff}=3$ & $N_c^{\rm eff}=5$  & 
$N_c^{\rm eff}=\infty$ \\ \hline
$a_1$  &0.986 & 1.04  & 1.08 &  1.15  \\
$a_2$  &0.25 &  0.058 & -0.095 &  -0.325 \\
$a_3$  &$-13.9-22.6i$   & 61 & $120+18i$ &$211+45.3i$ \\
$a_4$  &$-344-113i$ & $-380-120i$  & $-410-127i$ & $-450-136i$ \\
$a_5$  &$-146-22.6i$ & $-52.7$  & $22+18i$ & $134+ 45.3i$ \\
$a_6$  &$- 493-113i$ & $-515-121i$  & $-530-127i$ & $-560-136i$ \\
$a_7$  &$ 0.04-2.73i$ & $-0.7-2.73i$  & $-1.24-2.73i$ & $-2.04-2.73i$  \\
$a_8$  &$2.98 -1.37i$ & $3.32-0.9i$  & $3.59-0.55i$ & 4   \\
$a_9$  &$-87.9- 2.73i$ & $-91.1-2.73i$  & $-93.7-2.73i$ & $-97.6-2.73i$ \\
$a_{10}$&$-29.3-1.37i$ & $-13-0.91i$  & $-0.04-0.55i$ & 19.48   \\
\hline
\end{tabular}}
\end{center}
\end{table}   

Before carrying out the phenomenological analysis, 
we would like to discuss the dynamical mechanism involved. 
We first come
to the QCD anomaly effect.
As pointed out in \cite{KP97,ALI98}, 
the QCD anamoly appears through 
the equation of motion
\be
\partial^\mu(\bar s\gamma_\mu\gamma_5 s)=2m_s\bar si\gamma_5 s+{\alpha_s
\over 4\pi}G_{\mu\nu}\tilde{G}^{\mu\nu}.
\en
Neglecting the $u$ and $d$ quark masses in the equations of motion
leads to 
\be
\la \eta' |{\alpha_s\over 4\pi}G\tilde G|0\ra=f_{\eta'}^u m^2_{\eta'} 
\en
and hence 
\be
\la\eta'|\bar s\gamma_5 s|0\ra=-i{m_{\eta'}^2\over
2m_s}\,\left(f_{\eta'}^s
-f^u_{\eta'}\right).
\en
 To determine the decay constant $f_{\eta'}^q$, we need to know the
wave functions of the physical $\eta'$ and $\eta$ states which are related
to
that of the SU(3) singlet state $\eta_0$ and octet state $\eta_8$ by
\be
\eta'=\eta_8\sin\theta+\eta_0\cos\theta, \qquad
\eta=\eta_8\cos\theta-\eta_0
\sin\theta,
\en
with 
$|\eta_0\ra={1\over\sqrt{3}}|\bar uu+\bar dd+\bar ss\ra, \quad
|\eta_8\ra={1\over\sqrt{6}}|\bar uu+\bar dd-2\bar ss\ra $ and
$\theta\approx -20^\circ$.
When the $\eta-\eta'$ mixing angle is
$-19.5^\circ$,
the $\eta'$ and $\eta$ wave functions have simple expressions
\cite{Chau1}:
\be
|\eta'\ra={1\over\sqrt{6}}|\bar uu+\bar dd+2\bar ss\ra, \qquad
|\eta\ra={1\over\sqrt{3}}|\bar uu+\bar dd-\bar ss\ra.
\en
At this specific mixing angle, $f_{\eta'}^u={1\over 2}f_{\eta'}^s$ in the   
SU(3) limit. Introducing the decay constants $f_8$ and $f_0$ by
\be
\la 0|A_\mu^0|\eta_0\ra=if_0 p_\mu, \qquad \la 0|A_\mu^8|\eta_8\ra=if_8
p_\mu
\en
then $f_{\eta'}^u$ and $f_{\eta'}^s$ are related to $f_8$ and $f_0$ by
\cite{Ball}
\be
f_{\eta'}^u={f_8\over\sqrt{6}}\sin\theta+{f_0\over\sqrt{3}}\cos\theta,
\qquad f_{\eta'}^s=-2{f_8\over\sqrt{6}}\sin\theta+{f_0\over\sqrt{3}}\cos
\theta.
\en
Likewise, for the $\eta$ meson
\be
f_{\eta}^u={f_8\over\sqrt{6}}\cos\theta-{f_0\over\sqrt{3}}\sin\theta,
\qquad
f_{\eta}^s=-2{f_8\over\sqrt{6}}\cos\theta-{f_0\over\sqrt{3}}\sin\theta .
\en
From a recent analysis of the data of $\eta, \eta' \to \gamma \gamma$
and $\eta , \eta' \to \pi \gamma \gamma$ \cite{VH98}, $f_{8(0)}$ and
$\theta$ have been determined to be \be
{f_8 \over f_\pi}=1.38\pm0.22,~~~~~{f_0 \over f_\pi}=1.06\pm0.03,
~~~~\theta=-(22.0\pm3.3)^0,
\en
 which lead to
\be
f_{\eta}^u=99 {\rm MeV},~~~f_{\eta}^s=-108{\rm 
MeV},~~~f_{\eta'}^u=47{\rm MeV},~~~ f_{\eta'}^s=131{\rm MeV}.
\en
For the $u $ and $d $ quarks involved, we follow \cite{CT98} to  use
\be
\la\eta'|\bar u\gamma_5u|0\ra=\la\eta'|\bar d\gamma_5d|0\ra=r_{\eta'}
\,\la\eta'|\bar s\gamma_5s|0\ra,
\en
with $r_{\eta^{(')}}$ being given by
\be
r_{\eta'}=\,{\sqrt{2f_0^2-f_8^2}\over\sqrt{2f_8^2-f_0^2}}\,{\cos\theta+
{1\over \sqrt{2}}\sin\theta\over \cos\theta-\sqrt{2}\sin\theta}, \non \\
r_{\eta}=-{1\over 2}\,{\sqrt{2f_0^2-f_8^2}\over\sqrt{2f_8^2-f_0^2}}\,
{\cos\theta-\sqrt{2}\sin\theta\over \cos\theta+{1\over\sqrt{2}}
\sin\theta}.
\en

We next discuss the  $c\bar c \to\eta^{(')}$ mechanism.
 This new internal $W$-emission contribution
will be  important when the  mixing angle involved is $V_{cb}V_{cs}^*$, 
which is as large as that of the penguin amplitude and 
yet its effective parameter $a_2^{\rm eff}$ is larger than that of 
penguin operators. The decay constant $f_{\eta'}^{(\bar cc)}$,
defined as 
$\la 0|\bar c\gamma_\mu\gamma_5c|\eta'\ra=if_{\eta'}^{(\bar
cc)} q_\mu$, has been determined from the theoretical calculations
\cite{HZ97,SZ98,AMT} and from the phenomenological analysis of the 
data of $J/\psi\to\eta_c\gamma$, $J/\psi\to\eta' \gamma$
and of the $\eta\gamma$ and $\eta'\gamma$ transition form factors
\cite{Ali97,FK97}.
In the presence of the charm content in the $\eta_0$, an additional mixing
angle $\theta_c$ is needed to be introduced:
\be
|\eta_0\ra &=& {1\over \sqrt{3}}\cos\theta_c|u\bar u+d\bar d+s\bar s\ra
+\sin\theta_c|c\bar c\ra,   \non \\
|\eta_c\ra &=& -{1\over \sqrt{3}}\sin\theta_c|u\bar u+d\bar d+s\bar s\ra
+\cos\theta_c|c\bar c\ra.
\en
Then $f_{\eta'}^c=\cos\theta\tan\theta_c f_{\eta_c}$ and
$f_\eta^c=-\sin\theta\tan\theta_cf_{\eta_c}$,
where the decay constant $f_{\eta_c}$ can be extracted from $\eta_c\to 
\gamma\gamma$, and $\theta_c$ from $J/\psi\to\eta_c\gamma$ and
$J/\psi\to\eta'\gamma$ \cite{Ali97}. In the present paper we shall use
\be
f_{\eta'}^c=-6\,{\rm MeV},  \qquad f_\eta^c=-\tan\theta f_{\eta'}^c=
-2.4\,{\rm MeV},
\en
for $\theta=-22^\circ$, which are very close to the values
\be
f_{\eta'}^c=-(6.3\pm 0.6)\,{\rm MeV},  \qquad f_\eta^c=-(2.4\pm 0.2)\,{\rm
MeV}
\en
obtained in \cite{FK97}.
 
In the following we will show the input parameters we used.
One of the important parameters is the running quark mass which appears
in the matrix elements of $(S-P)(S+P)$ penguin operators through the use
of equations of motion. 
The running quark mass should
be applied at the scale $\mu\sim m_b$ because the energy release in the
energetic two-body charmless decays of the $B$ meson is of order $m_b$.
In this paper, we use \cite{Fusaoku}
\be
&& m_u(m_b)=3.2\,{\rm MeV},  \qquad m_d(m_b)=6.4\,{\rm MeV},  \qquad
m_s(m_b)=105\,{\rm MeV},  \non \\
&&  m_c(m_b)=0.95\,{\rm GeV},  \qquad
m_b(m_b)=4.34\,{\rm GeV},
\en
in ensuing calculation, where we have applied $m_s=150$ MeV at $\mu=1$
GeV.
   
   It is convenient to parametrize the quark mixing matrix in terms of the
Wolfenstein parameters: $A,~\lambda,~\rho$ and $\eta$ , where
$A=0.81$ and $\lambda=0.22$ \cite{PDG96}.
 A recent analysis of all available experimental
constraints imposed on the Wolfenstein parameters yields \cite{Parodi}
\be
\bar{\rho}=\,0.156\pm 0.090\,,   \qquad \bar{\eta}=\,0.328\pm 0.054,
\en
where $\bar{\rho}=\rho(1-{\lambda^2\over 2})$ and
$\bar\eta=\eta(1-{\lambda
^2\over 2})$, and it implies that the negative $\rho$ region is excluded
at 93\% C.L.. In this paper, we  employ the  
representative values: $\rho=0.16$ and $\eta=0.34$, which satisfies the
constraint $\sqrt{\rho^2+\eta^2}=0.37$.

Under the factorization approach, 
the decay amplitudes are expressed as
the products of the decay constants and the form factors. We use
the standard parametrization for decay constants and form
factors \cite{BSW}.
For values of the decay constants, we use $f_\pi=132$ MeV, $f_ K=160$
MeV, $f_\rho=210$ MeV,  $f_{K^*}=221$ MeV,  $f_ \omega=195$ MeV and 
$f_\phi=237$ MeV.
Concerning the  heavy-to-light mesonic form factors, 
we will use the results evaluated in the relativistic quark
model \cite{BSW,CCW} by
directly calculating
$B_{s(u,d)} \to P$ and $B_{s(u,d)} \to V$ form factors at time-like 
momentum transfer. Denoting $\eta_s = \bar{s}s$, the explicit
values for the form factors involved are 
$F_{1,0}^{B_s \eta_{s}}(0)=0.48$,
$F_{(0,1)}^{B_s \eta'_{s}}(0)=0.44$,
and $A_0^{B_s K^*}(0)=0.28$,  
which are larger than  BSW model's results \cite{DX}. 
The $q^2$ dependence of the matrix element, 
parametrized under the pole dominance ansatz, are found to have a dipole
behaviour for $A_0$ ,$F_1$, and a monopole one  for $F_0$.   
In the following, we will use the exact value calculated at the relevant 
kinematical point in this paper.
Note that these matrix elements should be used with a 
correct normalization \cite{CT98}, for which to a good approximation, we
take
$F_{1,0}^{B_s \eta}(0)={-1\over \sqrt{3}}F_{1,0}^{B_s \eta_s}(0)$
and 
$F_{1,0}^{B_s \eta'}(0)={2\over \sqrt{6}}F_{1,0}^{B_s \eta'_s}(0)$.

\medskip
{\bf 3.}~{\it Phenomenology}~~~ 
We are now ready to discuss the phenomenology of  exclusive
charmless rare $B_s$ decays. To illustrate the issue of $N_c^{\rm eff}$ 
dependence ( which means the different factorization approach)  of 
theoretical predictions, we will begin with 
$\bar{B_s} \to (\pi,\rho,\omega) \eta^{(')}$.
Unlike the case of  $B \to (\pi,\rho)\eta^{(')}$, 
$\bar{B_s}\to (\pi , \rho , \omega) \eta^{(')}$ do 
not receive the anomaly contribution from the $(S-P)(S+P)$ 
penquin operators 
due to the particle content of $\bar{B_s}$  and $\pi(\rho ,\omega)$. 
The decay amplitude for $\bar{B_s}^0\to\pi \eta^{(')}$ reads
\be
A(\bar{B_s}\to\eta^{(')} \pi) = {G_F\over\sqrt{2}}\Bigg\{
V_{ub}V_{us}^* a_2 
- V_{tb}V_{ts}^*\Bigg[ {3\over 2}(-a_7+a_9)\Bigg]\Bigg\}
X^{(B_s\eta^{(')},\pi)}_u,   \label{eta'pi}
\en
where
\be
X^{(B_s\eta^{(')},\pi)} &\equiv& \la \pi^0|(\bar
uu)_\vma|0\ra\la\eta^{(')}|(\bar
sb)_\vma|\bar{B_s}\ra
= -i{{f_\pi} \over
\sqrt{2}}(m_{B_s}^2-m^2_{\eta^{(')}})F_0^{B_s\eta^{(')}}(m_\pi^2).
\en
Since the internal W-emission is CKM-suppressed and the QCD penguins are
canceled  out in these decay modes, $\bar{B_s}\to\pi(\rho)\eta^{(')}$ are
dominated
by the EW penguin diagram. The dominant EW penguin contribution 
proportional to $a_9$
is $N_c^{\rm eff}$-stable, whereas the internal $W$ contribution $a_2$ is 
$N_c^{\rm eff}$-sensitive. Within the ``heterogeneous" nonfactorizable 
picture, $a_2$ is fixed and thus the predicted branching ratio is rather 
stable under the 
variation of $N_c^{\rm eff}$ as shown in the last four columns in
Table II. 
However, $a_2$  varies within the ``homogeneous" nonfactorizable
scheme
and thus the predicted branching ratios do show a $N_c^{\rm eff}$ 
dependence.
We would like to emphasize that although $\bar{B_s}\to(\pi,\rho)\eta^{(')}$ 
are 
dominated by the EW penguin diagram, the internal W diagram does make some 
contributions to this decay mode. Since $a_2$  changes sign from 
$N_c^{\rm eff}=2,3$ to $N_c^{\rm eff}=5, \infty$, the interference 
pattern between the internal $W$ diagram and the EW penguin diagram 
will change from the destructive to the constructive one. 
It is thus easy to see that for $N_c^{\rm eff}=2$ there is a  
larger destructive interference between the internal $W$ diagram and 
the EW penguin contribution and the predicted branching ratio is 
the smallest one among the first four columns in Table II, whereas for
$N_c^{\rm eff}=\infty$ constructive interference takes the role and   
the branching ratio increases.  

While QCD penguin diagrams are canceled out in $\bar{B_s}\to 
(\pi , \rho)\eta^{(')}$,
$\bar{B_s}\to\omega\eta^{(')}$ gets enhanced from the QCD penguin 
diagram. The decay amplitude for $\bar{B_s}\to\omega\eta^{(')}$ is 
\be
A(\bar{B_s}\to \omega \eta^{(')} ) = {G_F\over\sqrt{2}}\Bigg\{
V_{ub}V_{us}^* a_2
- V_{tb}V_{ts}^*\Bigg[ 2(a_3+a_5)+{1\over 2}(a_7+a_9)\Bigg]\Bigg\}
X^{(B_s\eta^{(')},\omega)}_u,   \label{eta'omega}
\en
where
\be
X^{(B_s\eta^{(')},\omega)} &\equiv& \la \omega|(\bar
uu)_\vma|0\ra\la\eta^{(')}|(\bar
sb)_\vma|\bar{B_s}\ra
= {\sqrt{2}} {f_\omega} m_{\omega}F_1^{B_s\eta^{(')}}(m_\omega^2)(\vp\cdot
p_{B_s}).
\en
From Table II, we see that there is a wide range of predictions 
for the branching ratios.
This process is QCD penguin dominated, except when
the ``naive" factorization is used 
or $(N_c^{\rm eff}(V-A),N_c^{\rm eff}(V+A))=(2,5)$ where there are large 
cancellations 
between the QCD penguin contributions ( $i.e.$ $a_3$ + $a_5$). 
The largest branching ration predicted for $B_s\to\omega\eta^{(')}$ occurs 
when we use the ``large-$N_c$ improved" factorization, where the EW 
penguin and QCD penguin have constructive interference.

Next, we discuss
$\bar{B_s}^0 \to \eta' K^0$ decay, which has the decay amplitude 
\be
A(\bar{B_s}^0\to K^0 \eta^{'})
&=& {G_F\over\sqrt{2}}\Bigg\{
V_{ub}V_{ud}^*a_2X^{({B_s}K,\eta^{'})}_u
+V_{cb}V_{cd}^*a_2 X^{({B_s}K,\eta^{'})}_c \non \\
&-& V_{tb}V_{td}^*\Bigg[ \left(a_4-{1\over 2}a_{10} 
+(2a_6-a_8){m_K^2\over
(m_s+m_d)(m_b-m_d)}\right)X^{({B_s}\eta^{'},K)}   \non \\
&+& \left(a_3-a_5-a_7+a_9\right)X^{({B_s}K,\eta^{'})}_u   
+ (a_3-a_5+{1 \over 2}a_7-{1\over 2}a_9)X^{({B_s}K,\eta^{'})}_s \non \\
&+& (a_3-a_5-a_7+a_9)X_c^{({B_s}K,\eta^{'})}   \non \\
&+& \Bigg(a_3+a_4-a_5+{1\over 2}a_7-{1\over 2}a_9-{1\over 2}a_{10} \non\\
&+& (a_6-{1\over 2}a_8){m^2_{\eta^{'}}\over
m_s(m_b-m_s)}\left({{f_{\eta^{'}}^s
\over f_{\eta'}^u} -1}\right) 
r_{\eta^{'}}\Bigg)X^{({B_s}K,\eta^{'})}_d \Bigg]\Bigg\},
\en
 where
\be
X^{(B_s\eta^{'},K)}
&=& \la K^0|(\bar sd)_\vma|0\ra\la\eta^{'}|(\bar
db)_\vma|\bar B_s^0\ra
= -if_K(m_{B_s}^2-m^2_{\eta^{'}})F_0^{B_s\eta^{'}}(m_K^2),   \non \\
X^{(B_s K,\eta^{'})}_q
&=& \la \eta^{'}|(\bar qq)_\vma|0\ra \la K^0|(\bar sb)_
\vma|\bar B_s^0\ra
= -if_{\eta^{'}}^q(m_{B_s}^2-m^2_K)F_0^{B_s K}(m_{\eta^{'}}^2).
\en
Due to the QCD anamoly, there is an extra  
$(f_{\eta'}^s/f_{\eta'}^u -1)$ term multiplied with $a_6$  in
$X^{(B_s K,\eta')}_d$,  whose presence is 
necessary in order to be consistent with the chiral-limit behaviour  of
the (S-P)(S+P) penguin matrix elements \cite{CT98}. 
Though penguin diagrams play the dominant role, the internal 
$W$ diagram and the mechanism of the $c\bar c$ pair into 
the $\eta'$ do have some nonneglible effects when $N_c^{\rm eff}=2$ and 
$N_c^{\rm eff}=\infty$ where $a_2$ gets the larger values. 
Due to the large cancellation, 
the EW penguin has only tiny effect and can be 
neglected.
The monotonic decrease of the branching ratio from $N_c^{\rm eff} =\infty$ 
to $N_c^{\rm eff}=2$ within the ``homogeneous" nonfactorizable picture
can be 
understood from the behaviour of the QCD
penguin $i.e.$ the destructive interference between $a_{(3,5)}$ and 
$a_{(4,6)}$: as $N_c^{\rm eff}$ decreases, $a_{(3,5)}$ contributions
increase and hence the branching ratios decrease. As we already mentioned
before, 
$a_3$ and $a_5$ are $N_c^{\rm eff}$-sensitive while $a_3-a_5$
is  stable under the variation of  $N_c^{\rm eff}$ and then the 
predicted branching ratio is $N_c^{\rm eff}$-stable within the
``homogeneous" factorization approach. 

There exist some general rules for the derivation of the 
formula from $ B 
\to P_a P_b$ to its corresponding $ B \to V_a P_b $ and  
$ B \to P_a V_b$.
These general rules can be written as:
(i) For $X^{(B P_a,P_b)}$ to $X^{(B V_a,P_b)}$, replace
the term $m^2_{P_b}/[(m_1+m_2)(m_3-m_4)]$ by 
$-m^2_{P_b}/[(m_1+m_2)(m_3+m_4)]$ and the index $P_a$ by $V_a$,  
(ii) discard the $(S-P)(S+P)$ contribution associated with
$X^{(B_s P_a,V_b)}$ and $a_{(5,7)} \to -a_{(5,7)}$ if they contribute.
Thus, the factorizable amplitude of $\bar{B_s} \to \eta' K^*$
can be readily obtained from the $\bar{B_s} \to \eta' K$ one and reads
\be
A(\bar{B_s} \to\eta^{'} K^{*0}) &=& {G_F\over\sqrt{2}}\Bigg\{
V_{ub}V_{ud}^*\left(a_2X^{(B_s K^*,\eta^{'})}_u \right)
+V_{cb}V_{cd}^*a_2X^{(B_sK^*,\eta^{'})}_c \non \\
&-& V_{tb}V_{td}^*\Bigg[
(a_4-{1\over 2}a_{10})X^{(B_s\eta^{'},K^*)}
+ \Bigg(a_3+a_4-a_5+{1\over 2}a_7-{1\over 2}a_9-{1\over 2}a_{10} \non \\
&-& (a_6-{1\over 2}a_8){m^2_{\eta^{'}}\over
m_s(m_b+m_s)}
\left({{f_{\eta^{'}}^s \over f_{\eta^{'}}^u} -1}\right) r_{\eta^{'}} 
\Bigg)X^{(B_sK^*,\eta^{'})}_d \non \\  
&+& \left(a_3-a_5-a_7+a_9\right)X^{(B_sK^*,\eta^{(')})}_u
+ (a_3-a_5+{1\over 2}a_7-{1\over 2}a_9)X^{(B_sK^*,\eta^{(')})}_s
\non \\
&+&(a_3-a_5-a_7+a_9)X_c^{(B_s K^*,\eta^{(')})}
\Bigg]\Bigg\},
\en
with
\be   
X^{(B_s \eta^{(')},K^*)}   
&\equiv& \la K^{*-}|(\bar su)_\vma|0\ra\la\eta^{(')}|(\bar ub)_
\vma|\bar{B_s}\ra
= 2f_{K^*}m_{K^*}F_1^{B_s\eta^{(')}}(m_{K^*}^2)(\vp\cdot p_{B_s}),   
\non \\ X^{(B_s K^*,\eta^{(')})}_q
&\equiv& \la \eta^{(')}|(\bar qq)_\vma|0\ra\la K^{*-}|(\bar
sb)_\vma|\bar{B_s} \ra
= 2f_{\eta^{(')}}^q m_{K^*}A_0^{B_s K^*}(m_{\eta^{(')}}^2)(\vp\cdot 
p_{B_s}). \en
It is interesting to see that since there is no $a_6$ term
in $X^{(B_s K^*,\eta')}$, the penguin contribution is reduced 
substantially
and so does the branching ratio. With a reduced penguin contribution, 
the involved internal $W$ diagram and  the mechanism of the $c\bar c$ 
pair into the $\eta'$ become more important than those of 
the $\bar{B_s} \to K \eta'$.
The larger branching ratios in columns denoted by $I_a$ and $II_b$ 
are due to the 
constructive interference between the internal $W$ diagram and the penguin
contribution.

We are now coming to the most complicated process ${\bar B_s} \to \eta 
\eta'$ , which has the decay amplitude  
\be
A(\bar{B_s}\to\eta \eta')
&=& {G_F\over\sqrt{2}}\Bigg\{
V_{ub}V_{us}^*\left(a_2X^{(B_s\eta,\eta')}_u+a_2X^{(B_s\eta',\eta)}_u\right)
+V_{cb}V_{cs}^*\left(a_2X^{(B_s\eta,\eta')}_c+a_2X^{(B_s\eta',\eta)}_c\right)
\non \\
&-& V_{tb}V_{ts}^*\Bigg[
 \Bigg(a_3+a_4-a_5+{1\over 2}a_7-{1\over 2}a_9-{1\over 2}a_{10} \non \\
&+&(a_6-{1\over 2}a_8){m^2_{\eta'}\over 
m_s(m_b-m_s)}\left(1-{f_{\eta'}^u
\over f_{\eta'}^s}\right)\Bigg)X^{(B_s\eta,\eta')}_s   \non \\
&+& \left(2a_3-2a_5-{1\over 2}a_7+{1 
\over 2}a_9\right)X^{(B_s\eta,\eta')}_u
+(a_3-a_5-a_7+a_9)X_c^{(B_s\eta,\eta')} \non \\
&+& \Bigg(a_3+a_4-a_5+{1\over 2}a_7-{1\over 2}a_9-{1\over 2}a_{10} \non \\
&+& (a_6-{1\over 2}a_8){m^2_{\eta}\over 
m_s(m_b-m_s)}\left(1-{f_{\eta}^u   
\over f_{\eta}^s}\right)\Bigg)X^{(B_s\eta',\eta)}_s   \non \\
&+& \left(2a_3-2a_5-{1\over 
2}a_7+{1\over 2}a_9\right)X^{(B_s\eta',\eta)}_u
+(a_3-a_5-a_7+a_9)X_c^{(B_s\eta',\eta)}
\Bigg]\Bigg\},
\en
with
\be
X^{(B_s\eta,\eta')}_q   
&\equiv& \la \eta'|(\bar qq)_\vma|0\ra\la
\eta|(\bar sb)_\vma|\bar{B_s}\ra
= -if_{\eta'}^q(m_{B_s}^2-m_{\eta}^2)F_0^{B_s \eta}(m_{\eta'}^2),\non \\
X^{(B_s\eta',\eta)}_q
&\equiv& \la \eta|(\bar qq)_\vma|0\ra\la
\eta' |(\bar sb)_\vma|\bar{B_s}\ra
= -if_{\eta}^q(m_{B_s}^2-m_{\eta}^2)F_0^{B_s \eta'}(m_{\eta'}^2). \non
\en
The destructive interference between $X_{q=(u,c)}^{(B_s\eta,\eta')}$
and $X_{q=(u,c)}^{(B_s\eta',\eta)}$ makes the internal $W$-emission 
, $c {\bar c} \to \eta^{(')}$ and the corresponding penguin 
contributions smaller. 
The EW penguin is smaller than the QCD penguin by an order of magnitude
at the amplitude level and hence can be neglected.                  
The dominant QCD penguin contributions are governed
by  $X_{s}^{(B_s\eta,\eta')}$
and $X_{s}^{(B_s\eta',\eta)}$, which have the constructive interference.
The ($a_3-a_5$) term is positive, 
contrary to the negative $a_4$ and $a_6$ terms, 
and becomes smaller when $N_c^{\rm eff}$ 
increases within the  ``homogeneous" nonfactorizable structure.
Thus a monotonic increase of the branching ratio when $N_c^{\rm eff}$ 
increases comes mainly from this reduced destructive interference, 
within  the ``homogeneous" nonfactorizable structure.
Similar arguments are also applied to the
``heterogeneous" structure.

With the general rules (i) and (ii) mentioned before,
the decay amplitude for ${\bar B_s} \to \phi \eta^{(')}$ can be easily
obtained  from ${\bar B_s} \to \eta \eta'$:  
\be
A(\bar{B_s}\to\phi \eta^{(')})
&=& {G_F\over\sqrt{2}}\Bigg\{
V_{ub}V_{us}^* a_2X^{(B_s\phi,\eta^{(')})}_u
+V_{cb}V_{cs}^*a_2X^{(B_s\phi,\eta^{(')})}_c \non \\
&-& V_{tb}V_{ts}^*\Bigg[
 \Bigg(a_3+a_4-a_5+{1\over 2}a_7-{1\over 2}a_9-{1\over 2}a_{10} \non \\
&-&(a_6-{1\over 2}a_8){m^2_{\eta^{(')}}\over
m_s(m_b+m_s)}\left(1-{f_{\eta^{(')}}^u
\over f_{\eta^{(')}}^s}\right)\Bigg)X^{(B_s\phi,\eta^{(')})}_s   \non \\
&+& \left(2a_3-2a_5-{1\over 2}a_7+{1\over 
2}a_9\right)X^{(B_s\phi,\eta^{(')})}_u
+(a_3-a_5-a_7+a_9)X_c^{(B_s\phi,\eta^{(')})} \non \\
&+& \Bigg(a_3+a_4+a_5-{1\over 2}a_7-{1\over 2}a_9-{1\over 2}a_{10} 
\Bigg)X^{(B_s\eta^{(')},\phi)}
\Bigg]\Bigg\},
\en
with
\be
X^{(B_s\phi,\eta^{(')})}_q
&\equiv& \la \eta^{(')}|(\bar qq)_\vma|0\ra\la
\phi|(\bar sb)_\vma|\bar{B_s}\ra  
= 2f_{\eta'}^qm_{\phi}A_0^{B_s \phi}(m_{\eta^{(')}}^2)(\vp\cdot p_{B_s}),     
\non \\
X^{(B_s\eta^{(')},\phi)}_q
&\equiv& \la \phi|(\bar qq)_\vma|0\ra\la
\eta^{(')} |(\bar sb)_\vma|\bar{B_s}\ra
= 2f_{\phi} m_{\phi}F_1^{B_s \eta^{(')}}(m_{\phi}^2)(\vp\cdot p_{B_s}). 
\non
\en
While the internal $W$ diagram is subject to the CKM-suppression,
the  $c {\bar c} \to \eta^{(')}$ mechanism  suffers from
the suppression in the decay constant and thus 
${\bar B_s} \to \phi \eta^{(')}$ is  dominated by the penguin 
contribution. Due to the cancellation among the different $a_i(i=3,4,5,6)$'s, 
effect of the QCD penguin, though still dominant, are reduced
substantially.
Within the ``homogeneous" nonfactorizable picture, we find 
a  monotonic decrease of the branching ratios when $N_c^{\rm eff}$ 
increases, which comes from a monotonic decrease 
of the QCD penguin contributions as $N_c^{\rm eff}$ increases. 
Since the QCD penguin contributions are reduced, the
EW penguin contributions become important. 
It is found that the interference pattern 
between the QCD  and EW penguin is destructive except for 
the ``large-$N_c$ improved" factorization approach, where
a constructive interference exists and a very 
dramatically suppressed QCD penguin contribution appears.
The strength of the destructive interference depends on  
$N_c^{\rm eff}$ and its effect is to reduce the QCD penguin
contribution without changing the trend of reduced penguin 
when $N_c^{\rm eff}$ increases.
A dramatic destructive interference among the penguin contributions occurs 
for ${\bar B_s} \to \phi \eta^{(')}$ when 
$N_c^{\rm eff}(V-A)=N_c^{\rm eff}(V+A)=\infty$ and 
$(N_c^{\rm eff}(V-A),N_c^{\rm eff}(V+A))=(2,\infty)$, thus the 
internal $W$ diagram and the $c {\bar c} \to \eta^{(')}$ mechanism
contribution becomes important relative to the other cases and the branching 
ratio is the smallest in these situations.

\vskip 0.4cm
\begin{table}
{{\small Table II. Average branching ratios (in units of $10^{-6}$) for 
charmless $B_s$ decays to $\eta'$ and $\eta$.
Predictions are for $k^2=m_b^2/2$, $\eta=0.34,~\rho=0.16$. 
I denotes the ``homogeneous" nonfactorizable contributions
{\rm i.e. $N_c^{\rm eff}(V-A)=N_c^{\rm eff}(V+A)$} and (a,b,c,d) represent 
the cases for $N_c^{\rm eff}$=($\infty$,5,3,2).
II denotes the ``heterogeneous" nonfactorizable contributions, 
{\rm i.e. $N_c^{\rm eff}(V-A) \neq N_c^{\rm eff}(V+A)$}
} and ($a'$,$b'$,$c'$) represent the cases for 
$N_c^{\rm eff}(V+A)$=(3,5,$\infty$), 
where we have fixed $N_c^{\rm eff}(V-A)$=2 (see 
the text) }

{\footnotesize
\begin{center}
\begin{tabular}{|l|c c c c |c c c |} \hline
Decay & $I_a$ & $I_b$ & $I_c$ & $I_d$ &
$II_{a'}$ & $II_{b'}$ & $II_{c'}$ \\ 
\hline 
$ \bar{B_s}\to\pi \eta' $ & 0.25 & 0.17 &0.13&0.11 &0.11&0.11&0.10 \\
$ \bar{B_s}\to\pi \eta $  & 0.16 & 0.11& 0.08 & 0.07 & 0.07&0.068&0.067 \\
$ \bar{B_s}\to \rho \eta' $ & 0.70&0.47&0.36&0.30&0.30&0.30&0.31 \\
$ \bar{B_s}\to\rho \eta $ & 0.45&0.30&0.24&0.19&0.19&0.19&0.20 \\
$ \bar{B_s}\to \omega \eta' $ & 6.9&0.9&0.012&2.14&0.48&0.03&0.83 \\
$ \bar{B_s}\to \omega \eta $ &4.45 &0.63&0.008&1.39&0.31&0.02&0.54 \\
$ \bar{B_s}\to\eta' K^0$ &1.25&1.07&1.01&1.00&1.27&1.51&1.90 \\
$ \bar{B_s}\to\eta K^0$ &1.35 &0.81&0.68&0.76&0.75 &0.74&0.72\\
$ \bar{B_s}\to \eta' K^{*0}$ & 0.49 & 0.35 &0.32 &0.26 & 0.49&0.60&0.80\\ 
$ \bar{B_s}\to\eta K^{*0}$ &0.45 & 0.05  & 0.02  & 0.24 & 0.24&0.24&0.25\\ 
$ \bar{B_s}\to\eta \eta'$ & 47.4&41.8 &38.3 &34.4 &39.5&44.1&51.5 \\
$ \bar{B_s}\to\eta' \eta'$ & 26.6 & 24.9&23.8 &22.4&33.8&43.9&62.2 \\
$ \bar{B_s}\to\eta \eta$ &20.3  &17.1 &15.1 &12.8&11.6&10.7&9.1 \\
$ \bar{B_s}\to\phi \eta'$ & 0.44 &0.59 &2.29 &6.20&4.41&3.11&1.66 \\
$ \bar{B_s}\to\phi \eta$ & 0.04 &0.91 &2.29 &4.92&2.28 &0.92&0.10   \\
\hline
\end{tabular}
\end{center} } 
\end{table} 
\vskip 0.4cm

{\bf 4.}~{\it Summary and Discussions}~~To summarize, we have studied 
charmless exclusive nonleptonic $B_s$ meson decay into an $\eta$
or $\eta'$  within the generalized factorization approach.
Nonfactorizable contributions are parametrized in terms of  the 
effective number of colors 
$N_c^{\rm eff}$ and predictions using different factorization approaches 
are shown with the $N_c^{\rm eff}$ dependence. 
It is found that for processes depending on the $N_c^{\rm eff}$-stable 
$a_i$'s such as
$\bar{B_s} \to (\pi , \rho) \eta^{(')}$, the branching ratios are not
sensitive 
to the factorization approach we used. 
While for the processes depending on  the $N_c^{\rm eff}$-sensitive
$a_i$'s such as the  
$\bar{B_s} \to \omega \eta^{(')}$, the predicted
branching ratios have a wide range
depending on the choice of the factorization approach. 
The effect of the QCD anomaly, which 
is not discussed in the earlier literature, 
is found to be important for the $\bar{B_s} \to \eta^{(')} \eta^{(')}$. 
We also found 
that the mechanism $(c\bar c) \to \eta'$
, in general, has smaller effects  due to a
possible CKM-suppression and the suppression in the  decay constants
except for the $\bar{B_s} \to \phi \eta$ 
under the ``large-$N_c$ improved" factorization approach,
where the internal $W$ diagram is CKM-suppressed 
and the penguin contributions are compensated.

In this Letter, we, following the 
standard approach, have neglected the $W$-exchange and the
space-like penguin contributions. 
Another major source of uncertainties comes from the
form factors we used, which are larger than the BSW model's calculations.
Although the Wolfenstein parameter $\rho$ ranges from the negative region 
to the positive one, we have ``fixed" it to some representative values. 
The interference pattern between the internal $W$ diagram and the penguin 
contributions will change when we take a different sign of $\rho$.
We will study these form factor- and CKM- dependence involved and all 
the $B_s \to PP,VP,VV$ decay modes in a separate publication. 

\bigskip
\noindent ACKNOWLEDGMENT:~~We are very grateful to Prof. H.Y. Cheng for 
helpful discussions.
This work is supported in part by the National Science Council of
the Republic of China under Grant NSC87-2112-M006-018.

\renewcommand{\baselinestretch}{1.1}
\newcommand{\bi}{\bibitem}

\newpage

\end{document}